\documentclass[]{raa}            % referee version: for submission

\usepackage{graphicx,times}             %for PS/EPS graphics inclusion, new
\usepackage{csvsimple}
\usepackage{amssymb }
\usepackage{natbib  }

\begin{document}

   \title{Database of Molecular Masers and Variable Stars
%\footnotetext{$*$ Supported by the National Natural Science Foundation of China.}
}
%   \subtitle{I. Place Your Subtitle Here}

   \volnopage{Vol.0 (20xx) No.0, 000--000}      %%preserved for Editor. DOn't remove!
   \setcounter{page}{1}          %%starting page, preserved for Editor. DOn't remove!

   \author{Andrej M. Sobolev
      \inst{1}
   \and Dmitry A. Ladeyschikov
      \inst{1}
    \and Jun-ichi Nakashima
      \inst{2}
   }

   \institute{Astronomical Observatory, Ural Federal University, Lenin Avenue 51, 620000, Ekaterinburg, Russia; {\it Andrej.Sobolev@urfu.ru }\\
        \and
             Department of Astronomy, Geodesy and Environment Monitoring, Ural Federal University, Lenin Avenue 51, 620000, Ekaterinburg, Russia\\
   }

   \date{Received~~2018 month day; accepted~~2018~~month day}

%%% -------------------------------------------------------------------------------------------------

\abstract{We present the database of maser sources in H$_2$O, OH and SiO lines that can be used to identify and study variable stars at evolved stages. Detecting the maser emission in H$_2$O, OH and SiO molecules toward infrared-excess objects is one of the methods of identification long-period variables (LPVs, including Miras and Semi-Regular), because these stars exhibit maser activity in their circumstellar shells. Our sample contains 1803 known LPV objects. 46\% of these stars (832 objects) manifest maser emission in the line of at least one molecule: H$_2$O, OH or SiO. We use the database of circumstellar masers in order to search for long-periodic variables which are not included in the General Catalogue of Variable Stars (GCVS). Our database contains 4806 objects (3866 objects without associations in GCVS catalog) with maser detection in at least one molecule. Therefore it is possible to use the database in order to locate and study the large sample of long-period variable stars. Entry to the database at http://maserdb.net .
\keywords{catalogs --- Astronomical Databases: 
stars: variables: general --- Stars:
masers --- Physical Data and Processes}
}

   \authorrunning{A. M. Sobolev, D. A. Ladeyschikov \& J. Nakashima }            %author_head in even pages
   \titlerunning{Database of Molecular Masers and Variable Stars}  % title_head in odd pages

   \maketitle
%% The author head (on even pages) and the title head (on odd pages) will be
%% automatically extracted from \author{} and \title{}. Whenever the title is too long,
%% you will be asked to supply a shorter one by inserting either \authorrunning{} or
%% \titlerunning{} before \maketitle. Anyway, you can specify your own heads.
%%
%%
%% Note: In the following text body of your manuscript, please note several differences from
%%       other major journals:
%% (1) \subsection{Please Capitalize the First Letter of Each Notional Word in Subsection Title}
%% (2) Please Capitalize the First Letter of Each Notional Word in all tables' captions

%
%________________________________________________ sections below
%
\section{Introduction}           %% first-level sections will be auto-capitalized
\label{sect:intro}

At the end of their evolution, the stars of a few solar masses become red giants and increase the radius of the stellar atmosphere from 1~$R_\odot$ to 10$^2~R_\odot$. After that, the atmosphere is expelled into outer space and the star becomes a white dwarf surrounded by a planetary nebula. The red giant phase is usually short in time and lasts only a few hundred thousand years.

Many of red giant stars exhibit variability on the scales of a few hundred days. The variability of these stars is therefore called long-periodic, introducing long-period variables (LPVs). According to \cite{KHO85} LPVs are divided into two subgroups - Miras (M) with period usually between 150 and 600 days and magnitude variation $\Delta m>2.5$ and semiregular variables (SR) with a shorter period (50 to 150 days) and smaller magnitude variations ($\Delta m<2.5$). While leaving the star the matter forms a gas-dust circumstellar envelope. Large infrared excess is characteristic of these stars because the circumstellar envelope is re-emitting stellar light. 

The envelope of red giant stars manifests maser and thermal emission in a number of molecular lines. Maser emission arises in the lines of following molecules: OH ($\lambda$=18~cm), H$_2$O ($\lambda$=1.35~cm), SiO ($\lambda$=7~mm, 3.5~mm and others) and HCN ($\lambda$=3.3~mm). Thermal emission is found in the lines of CO ($\lambda$=2.6~mm), SiO, HCN and other molecules. Most of H$_2$O, OH and SiO masers are found in oxygen-rich M-type stars, but SiO masers are sometimes found in SR stars. HCN maser emission is found in carbon stars. Infrared emission of the star is responsible for pumping of the OH masers \citep{He05} and the impact of the shock waves onto the interior layers of circumstellar envelope is responsible for pumping of the SiO, H$_2$O and HCN masers \citep{Gray}. 

OH masers are most important for studies of Mira-type variables. Most of the LPVs that exhibit maser emission in OH lines have strong 1612 MHz emission. This is distinction from the OH masers in star-forming regions which have 1665 and 1667 MHz maser emission stronger than 1612 MHz one.

Until now, the compilations comprehensively including three major maser species in evolved stars (i.e., SiO, H$_2$O, OH) has been practically limited only to the catalog \cite{BEN90} which was published more than a quarter of century ago. For OH masers solely, there exists the University of Hamburg (UH) database, but there is no updated compilation work for H$_2$O and SiO masers. In order to utilize the information on masers in astrophysical reasearch it is highly desirable to have a database containing information on all three maser species. We are currently compiling a database including SiO, H$_2$O and OH masers (Nakashima et al., 2018, accepted to publication in IAU Symposium No. 336). This database consists of a web-service which realizes access to compiled maser observations from published papers and combines them with the newly collected data. Infrared and other data on each source is attached from the published surveys and catalogs. The archives currently used are the OH maser archive from \cite{ENG15}, and H$_2$O and SiO archives, which are currently under construction. So far, the information of about 27,000 observations ($\sim$11,000 objects) has been implemented. We also have a plan to extend the database by including methanol maser emission and other types of objects, such as young stellar objects, in future. 

%% Authors can give a citation as 'Michel et al. 1992'.
%% You may also use \cite, \citep and \citet for citation, and use Table~1 or Figure~1
%% and so forth. Using \ref and \label for cross-references of Tables/Figures
%% is a good way in adjusting/adding/removing text, tables or figures.

\section{Summary of the Collected Data}
\label{sect:data}

The initial release of the database is dedicated to the circumstellar maser sources of variable stars mainly in the following maser lines: SiO $J=1-0$, $v=1$ \& 2 (43 GHz), H$_2$O 22 GHz, OH 1612, 1665, 1667 MHz. The data are taken mainly from 5 published/unpublished compilation catalogs. The OH data are based on the OH maser archive from \cite{ENG15}. The H$_2$O data are based on an on-going compilation work (PI: Engels, D.). A significant amount of additional data of other maser transitions (for example, SiO $J=1-0$ $v=$0 \& 3, SiO $J=2-1$, $v=$1 \& 2, $^{29}$SiO $J=1-0$ $v=0$, etc.) are also included in the database, but the data survey for these lines are still not completed (the data will keep updating). We note that a non-negligible number of unpublished data of the Nobeyama SiO maser survey project are released to the public for the first time (the number of unpublished Nobeyama observations is about 400). In addition to the basic line parameters (such as intensity, velocity, line-profile, etc.), for a part of the observations, spectral data in ascii format are available, so that users could process the spectral data for their own purposes. In total, at this moment, ~11000 objects, which have been observed, at least, in one of the OH, H$_2$O or SiO maser lines or in the multiple maser lines, are included in the database (the distribution of the objects in the Galactic coordinates is given in Figure~1). Among the $\sim$11000 sources, the number of objects observed in the SiO, H$_2$O and OH maser lines are $\sim$4100, $\sim$4000, and $\sim$6700 respectively (overlaps exist between different maser species).

\begin{figure}[b]
% \vspace*{-2.0 cm}
\begin{center}
  \includegraphics[width=5in]{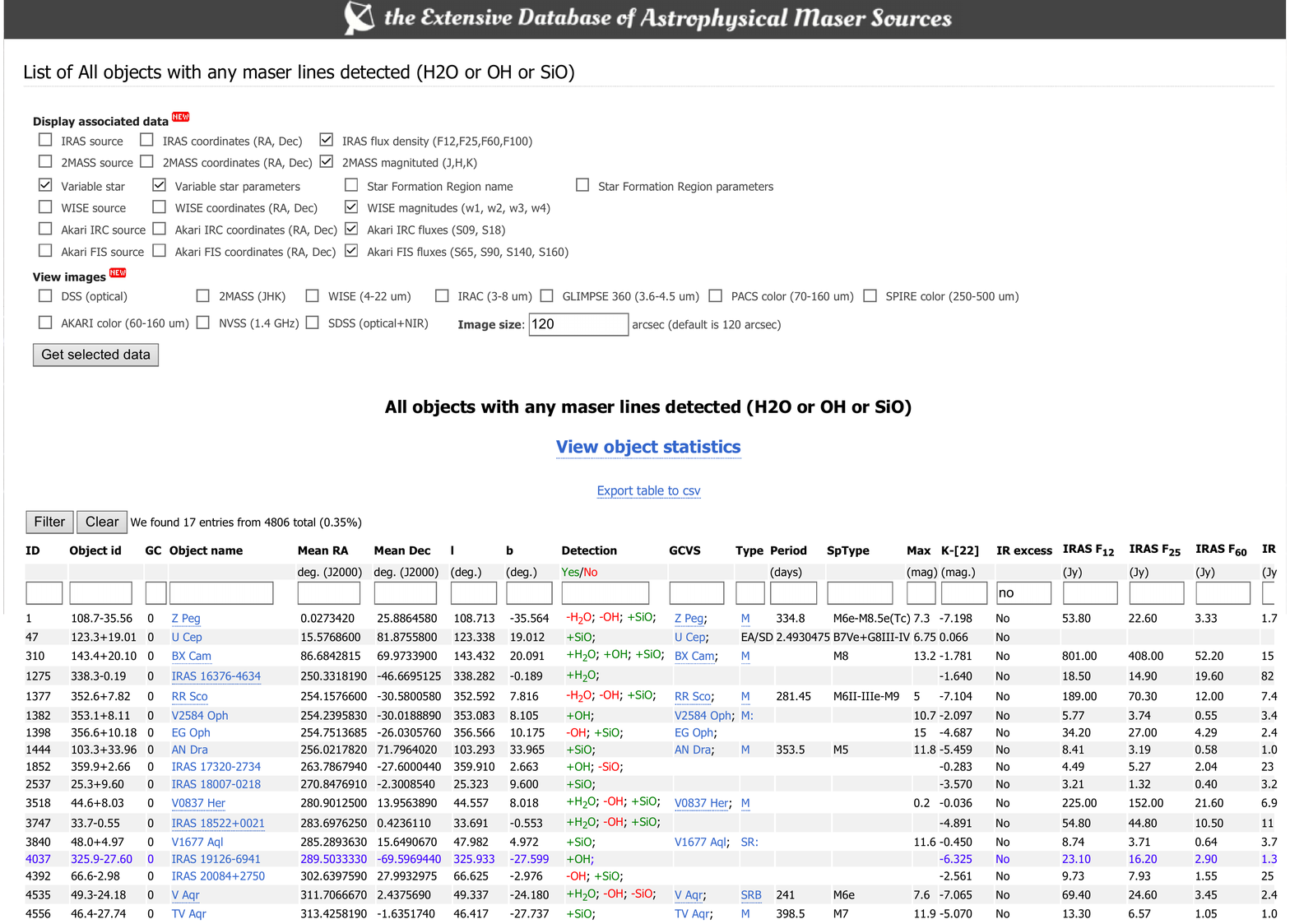} 
% \vspace*{-1.0 cm}
 \caption{List of objects from the web-based system displaying the maser detection in different molecules together with infrared characteristics.}
   \label{fig1}
\end{center}
\end{figure}

\section{Web-based system of the maser database}
\label{sect:data}
The web-based system\footnote{http://maserdb.net} for the database of the maser sources was developed in order to collect and display the large dataset of the maser sources. The system is written in the \textsc{Perl/CGI} language and use the external modules for full functionality, including  \textsc{Aladin Lite API}, \textsc{Vizier TAP} service, \textsc{Simbad} identification service and \textsc{ADS} service for publication search. The system allows to collect, display and analyze the large maser dataset from the available literature. Entering data to the database is done by using the CSV format that comes from optical character recognition of article text in PDF format or imported from \textsc{Vizier} archive if available. Entering the spectra profiles is done by using digitizing software -- \textsc{im2graph}. 

The list of the features of the web-system include the following:
\begin{itemize}
\item Search for maser data by coordinates, source name  or list of sources.
\item Parallel data search in popular astronomical catalogs from VizieR.
\item Association of maser observations with popular infrared and stellar catalogs - IRAS, 2MASS, UKIDSS, WISE, Akari, GCVS, etc. with instantaneous output of photometric and other data from these catalogs.
\item Cross-identification of masers in different molecules. The possibility of identifying objects in which emission is present in several maser molecules.
\item Ability to download the observational data in the CSV format.
\item Detailed research of each object in the database using images in different spectral ranges (from optical to radio).
\item For some observations ($\sim$3.2 thousand) there are spectra themselves, which can be viewed and analyzed directly in the system.
\item Statistical analysis of data - the construction of color-color diagrams, longitude-velocity, histograms of the spatial distribution of masers, etc. Ability to plot 1D histograms, 2D and 3D  distribution plots of any parameter sets. 
\end{itemize}

Development of the web-based system is not limited only to H$_2$O, OH and SiO molecules. The ability of input the CH$_3$OH (methanol) data is also included to the presented maser database system, allowing to study not only late-type stars (long-periodic variables, asymptotic giant branch stars) but also early-type objects (YSO, star formation regions, etc.).

\begin{figure}[b]
% \vspace*{-2.0 cm}
\begin{center}
 \includegraphics[width=5.5in]{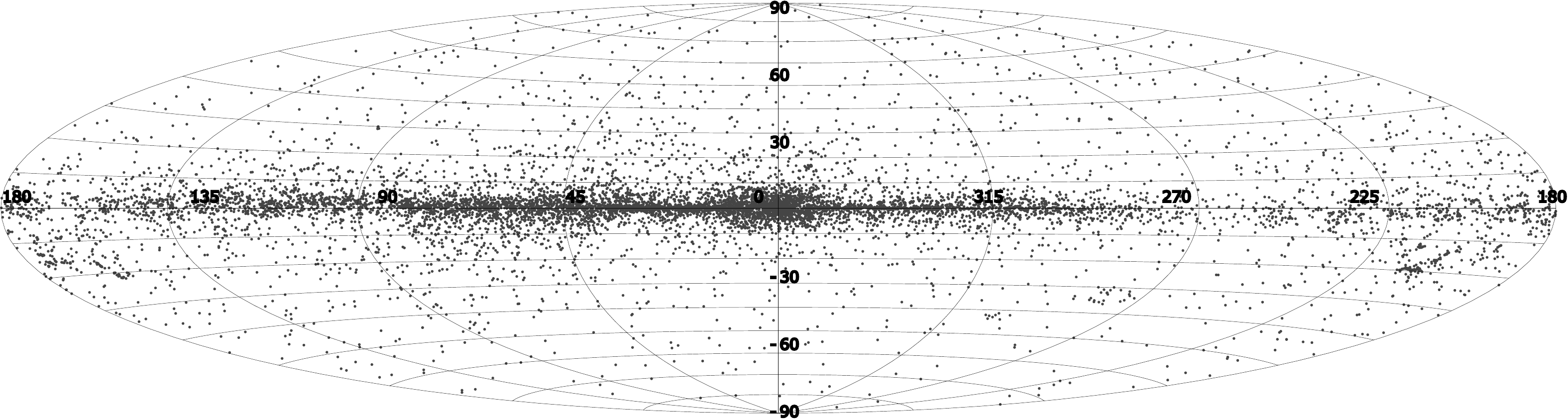} 
% \vspace*{-1.0 cm}
 \caption{Distribution of $\sim$11000 objects included in the database at the Galactic plane. These objects are observed, at least, in one of the OH, H$_2$O and SiO maser lines, and the both detection's and non-detection's are included.}
   \label{fig1}
\end{center}
\end{figure}

\begin{table}
\bc
\begin{minipage}[]{100mm}
\caption[]{Example of objects with maser emission that was not included to the GCVS catalog -- possible candidates for LPV's. Full version of this table is avaliable online at http://maserdb.net.\label{tab1}}
   \label{tab1}
\end{minipage}
\setlength{\tabcolsep}{1pt}
\small
 \begin{tabular}{cccccccccc}
  \hline\noalign{\smallskip}
     id & Galactic id &  Object name &  Mean RA  &  Mean Dec      &  Maser detection &  2MASS K &  WISE [22] &  K-[22] \\
           &          &                            &  Deg. (J2000)  &  Deg. (J2000)&  H$_2$O/OH/SiO  &  (mag.)    &  (mag.)      &  (mag.)   \\
  \hline
      \begin{filecontents*}{candidates_short.csv}
id,Galactic id,Object name,Mean RA,Mean Dec,Mean l,Mean b,Detection,F12,F25,F60,F100,2MASS J,2MASS H,2MASS K,W1mag,W2mag,W3mag,W4mag,K-[22],IR excess,References,References_short
1,117.7-7.60,IRAS 00127+5437,3.850714,54.9041685,117.721,-7.603,-H$_2$O~-OH~+SiO,56.00,39.6,5.48,6.55,6.793,5.23,4.159,3.81,1.531,0.128,-1.22,5.379,Yes,JIA96; SIV90; TAK01; YOO14;,[1][2][3][4]
2,119.7+3.32,IRAS 00170+6542,4.96446,65.991863,119.723,3.317,+H$_2$O~+OH~+SiO,27.20,32.5,6.89,21.5,10.147,7.798,6.047,3.55,1.001,0.056,-1.466,7.513,Yes,FUJ01; LIN89; LSQ92; WOL12; YUN13;,[5][6][7][8][9]
3,301.4-29.42,IRAS 03074-8732,43.096667,-87.338889,301.394,-29.419,+H$_2$O~+OH~+SiO,89.60,86.5,16.2,4.59,5.575,3.903,2.897,1.395,-0.062,-0.668,-2.391,5.288,Yes,DEG01; DEG89; LIN91A;,[10][11][12]
4,138.0+7.26,IRAS 03206+6521,51.285417,65.535279,137.97,7.256,-H$_2$O~+OH~+SiO,95.80,134,37.5,9.9,16.775,12.174,8.548,4.186,0.443,-0.341,-2.54,11.088,Yes,CHO12; TAK01; YUN13; ZUC87; DAV93B; ENG07; LIN89; LIN91A; LSQ92; WOL12; JEW91; JIA96; KIM14; NRO\_list;,[13][3][9][14][15][16][6][12][7][8][17][1][18][19]
5,141.7+3.53,IRAS 03293+6010,53.3775075,60.33597,141.725,3.519,-H$_2$O~+OH~+SiO,37.20,69.5,17.4,5.5,13.611,10.105,7.559,4.602,1.675,-0.209,-2.347,9.906,Yes,BEN96; KIM10; TAK01; DIC91; JEW91; NRO\_list;,[20][21][3][22][17][19]
6,161.1-17.17,IRAS 03453+3207,57.134583,32.278611,161.084,-17.172,+H$_2$O~-OH~+SiO,8.62,6.56,1.11,9.85,5.956,4.781,4.165,3.537,2.801,1.279,0.289,3.876,Yes,DEG12; LEW94; LEW95B; VAL01;,[23][24][25][26]
7,187.7-17.57,IRAS 04575+1251,75.099583,12.934903,187.712,-17.574,+H$_2$O~+OH~-SiO,133.00,122,19.6,6.95,8.735,6.425,4.861,0.845,-0.46,-0.686,-2.306,7.167,Yes,DAV93B; LEW94; LEW97; LSQ92; WOL12; NRO\_list; VAL01;,[15][24][27][7][8][19][26]
8,163.0+4.33,IRAS 05131+4530,79.19629,45.5677295,162.951,4.332,-H$_2$O~+OH~+SiO,27.80,48.8,14.9,3.12,12.186,8.69,6.331,3.291,0.19,-0.214,-2.307,8.638,Yes,BEN96; YUN13; DAV93B; LIN89; LIN91A; LSQ92; JIA99;,[20][9][15][6][12][7][28]
9,186.1-7.59,IRAS 05284+1945,82.852917,19.788611,186.056,-7.593,-H$_2$O~+OH~+SiO,4.17,8.98,3.35,1.5,18.396,16.036,11.643,6.548,4.155,1.349,-0.254,11.897,Yes,DAV93A; LEW94; LEW97; LIN91A; JIA96; YUN13;,[29][24][27][12][1][9]
10,176.5+0.18,IRAS 05345+3157,84.449259,31.99,176.517,0.182,+H$_2$O~+OH~-SiO,12.30,18.9,339,565,12.122,11.042,10.243,8.6,7.903,3.432,-0.345,10.588,Yes,JIA96; LEW93; SUN07;,[1][30][31]
11,174.1+14.12,IRAS 06297+4045,98.3155835,40.714139,174.114,14.122,+H$_2$O~+OH~+SiO,103.00,94.4,20.7,5.76,5.352,3.704,2.325,-0.64,-1.027,-1.489,-3.132,5.457,Yes,JEW91; KIM10; KIM14; NRO\_list; KIM10; VAL01; LIN89; LSQ92; WOL12;,[17][21][18][19][21][26][6][7][8]
12,215.5-6.08,IRAS 06319-0501,98.6091685,-5.060556,215.517,-6.08,+H$_2$O~+OH~-SiO,51.20,64.6,26.6,10,8.209,6.939,5.704,3.609,0.236,-0.191,-2.059,7.763,Yes,DEG07B; DEG89; KIM13; TAK01; VAL01; ZUC87; LEW95A; LIN91A;,[32][11][33][3][26][14][34][12]
13,212.1-4.07,IRAS 06329-0106,98.8829185,-1.156944,212.148,-4.066,-H$_2$O~-OH~+SiO,32.40,26.1,3.45,10.1,5.741,4.497,3.685,3.723,2.274,0.816,-0.338,4.023,Yes,CHO12; TAK01; DEG04B; DEG07B; JIA96; KIM14; LIN91A;,[13][3][35][32][1][18][12]
14,193.6+16.22,IRAS 07045+2418,106.893032,24.221952,192.666,14.133,+H$_2$O~+OH~+SiO,14.80,9.96,2.45,1,5.432,4.693,4.087,1.797,0.792,0.436,-0.477,4.564,Yes,DEG12; LEW94; LEW97; VAL01;,[23][24][27][26]
15,224.3-1.30,IRAS 07054-1039,106.95575,-10.73497,224.342,-1.288,+H$_2$O~+OH~+SiO,52.80,39.9,6.35,13,6.541,4.891,3.651,2.031,0.411,-0.27,-1.472,5.123,Yes,KIM10; KIM14; NRO\_list; LIN89; TAK94;,[21][18][19][6][36]
16,149.9+26.48,IRAS 07051+6601,107.5244695,65.940188,149.93,26.48,+H$_2$O~-OH~+SiO,52.90,31.9,5.12,1.71,5.321,4.224,3.29,1.569,-0.177,0.187,-0.86,4.15,Yes,DEG12; KIM14; KIM13; TAK01; LEW95A;,[23][18][33][3][34]
17,237.5-5.49,IRAS 07153-2411,109.365833,-24.286667,237.455,-5.487,+H$_2$O~-OH~+SiO,35.30,26.3,3.96,17.6,8.65,6.864,5.309,3.886,1.754,0.425,-0.8,6.109,Yes,CHO12; DEG04B; DEG07B; JIA96; KIM14; LIN91A;,[13][35][32][1][18][12]
18,228.1+0.21,IRAS 07180-1314,110.081458,-13.3369545,228.066,0.211,+H$_2$O~+OH~+SiO,40.70,46.7,7.8,1.83,6.515,4.724,3.534,1.224,-0.338,-0.535,-2.185,5.719,Yes,DEG04B; DEG07B; LEW95A; LSQ92; YOO14;,[35][32][34][7][4]
19,252.8-1.00,IRAS 08089-3511,122.6992825,-35.346389,252.813,-0.998,-H$_2$O~+OH~+SiO,38.80,35.1,6.75,13.1,6.971,5.305,4.242,3.825,1.561,0.384,-1.038,5.28,Yes,CHO12; TAK01; DEG04B; DEG07B; LIN91A;,[13][3][35][32][12]
20,235.3+18.10,IRAS 08357-1013,129.536602,-10.404549,235.323,18.098,+H$_2$O~+OH~+SiO,62.50,59.2,10.7,2.67,8.718,6.189,4.395,-0.121,-0.584,-0.932,-2.504,6.899,Yes,KIM10; KIM14; NRO\_list; LIN89; LSQ92; TAK01; TAK94; VAL01;,[21][18][19][6][7][3][36][26]
\end{filecontents*}
\csvreader[]{candidates_short.csv}{}
    {\\\csvcoli&\csvcolii&\csvcoliii&\csvcoliv&\csvcolv&\csvcolviii&\csvcolxv&\csvcolxix&\csvcolxx}

  {\smallskip}

\end{tabular}

\ec
%% place \tablecomments and \tablerefs below \end{center| and \end{center}:
%% you may leave the table-width parameter to editors or set to your actual size
%\tablecomments{0.86\textwidth}{ }
\end{table}

\section{Identification of variable stars with maser emission}
\label{sect:data}

Currently the General Catalogue of Variable Stars (GCVS; \citet{Samus17})  contains variable stars which are identified mostly by observations in the optics. Among them, Long-periodic stars (including Mira-type stars) are most common. The high detection rate of LPV stars is explained by the high luminosity of these stars (up to $10^3-10^4~L_{\odot}$) in the LPV stage and high amplitude of variability in visible light ($>2.5^m$ for Mira-type stars). LPV is a relatively short stage of the stellar evolution (only several hundred thousand years), thus they compose a relatively small portion of the total stellar population of the Galaxy. But this stage is very important. At this stage, all stars of low and intermediate mass are actively losing their matter, forming a gas-dust circumstellar shell and at the end forming a planetary nebula with the white dwarf in the center.

M-stars emit most of their light in the infrared because the circumstellar envelope heated by the starlight  is rather extended. Detection of Mira-type stars in the optics is sometimes missing. The optical light curve of some variable stars can have its minimum phase during the period of observations, thus these stars will be not detected in the optical range. Because variations in the magnitudes may be in the range from 2.5$^m$ to 11$^m$ and the period varies in the range 80-1000 days, only long-time monitoring observations (up to several years) may reveal variability of these stars. Infrared emission of these stars is more stable - variations of intensity in K band for M-type stars usually do not exceed 0.9$^m$ (see GCVS catalog for explanations).  

We are suggesting another way to identify candidates for LPV stars, including most common M-type stars. The maser emission in circumstellar shell of the late-type star has good prospects to identify and locate such stars. The envelope of red giant stars emits maser emission in the lines of H$_2$O, OH, SiO molecules and sometimes HCN (in Carbon-rich stars). We analyzed our database and considered all objects where maser emission of H$_2$O, OH and SiO molecules is detected. Because our database is cross-matched with GCVS catalog, we can explore maser detection rate for the stars of known type. 

At first, we select all objects from the database where at least one maser line is observed (H$_2$O, OH or SiO) in avaliable literature, including negative detections. We excluded 579 objects from the Galactic Center region ($359<l<1, -2<b<2$) because the high surface density of stars does not allow reliable association of maser observations with the known variable stars from GCVS catalog. There are total 10701 objects in the database, excluding Galactic Center region. Among them 2344 objects are included in the GCVS catalog. Then we select objects which are confirmed LPVs (Mira's and Semiregular variable stars) from GCVS catalog. We found 1803 such stars. Then we investigate maser detection in this LPV sample.

Among 1803 LPV stars 836 objects (46\%) have maser detection in at least one maser molecule (H$_2$O, OH or SiO). 331 objects (18\%) have maser detection in at least two maser molecules and 154 (8\%) have positive detection in all three maser molecules. Among 744 LPV stars with masers observations available in all three molecules (H2O, OH and SiO), including negative detections, only 30 stars have negative detection in all three maser molecules.  We can see that almost half of LPV stars exhibits maser emission in at least one maser line. 30 stars with negative detection in all three maser molecules may be associated with a minimum of maser variability. Therefore, high maser detection rate (46\%) for LPV stars shows that the maser emission has high potential in the  identification of LPV star candidates, including Mira's and Semiregular stars.

Table \ref{tab1} displays a list of objects which were selected using maser emission and are not included in  the GCVS catalog, along with their IR excess characteristics (see section~\ref{sect:data}).  These objects are potential candidates for Mira-type variable stars. In order to confirm the variability of these objects, the optical observations are needed. Alternatively, variability data can be taken from the new DR2 release of the GAIA catalogue \citep{Holl18}.

\section{Colour-colour diagrams} % IRAS, Akari, WISE

\begin{figure}
% \vspace*{-2.0 cm}
\begin{center}
 \includegraphics[width=4.5in]{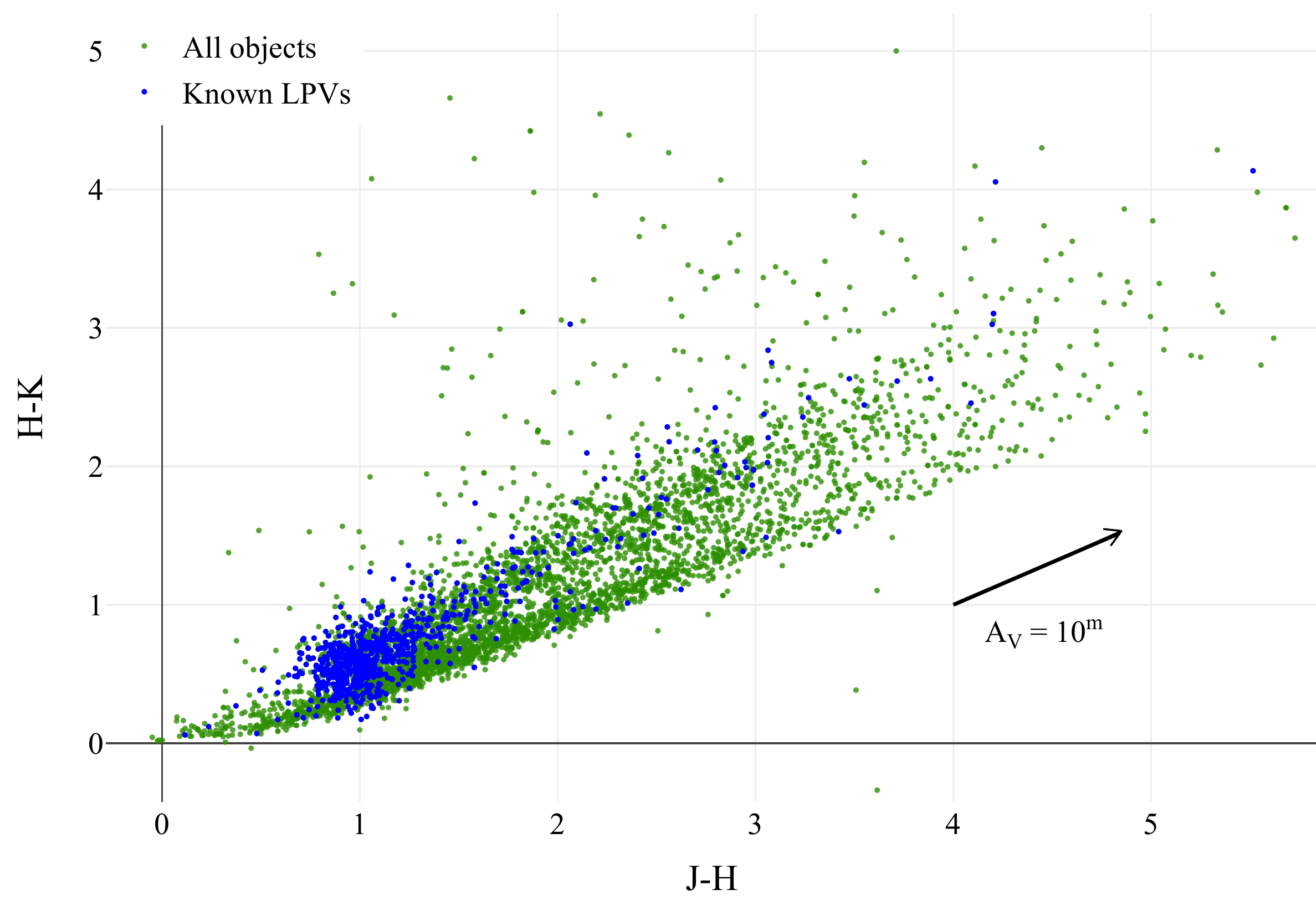}  
% \vspace*{-1.0 cm}
 \caption{2MASS colour-colour diagram for objects with detected maser emission of H2O, OH or SiO molecules. Blue dots represent known LPV stars (Miras and Semiregular variable stars) from GCVS catalog. Black line represent the reddening vector with A$_V=10^m$}
   \label{fig1}
\end{center}
\end{figure}

\begin{figure}
% \vspace*{-2.0 cm}
\begin{center}
 \includegraphics[width=4.5in]{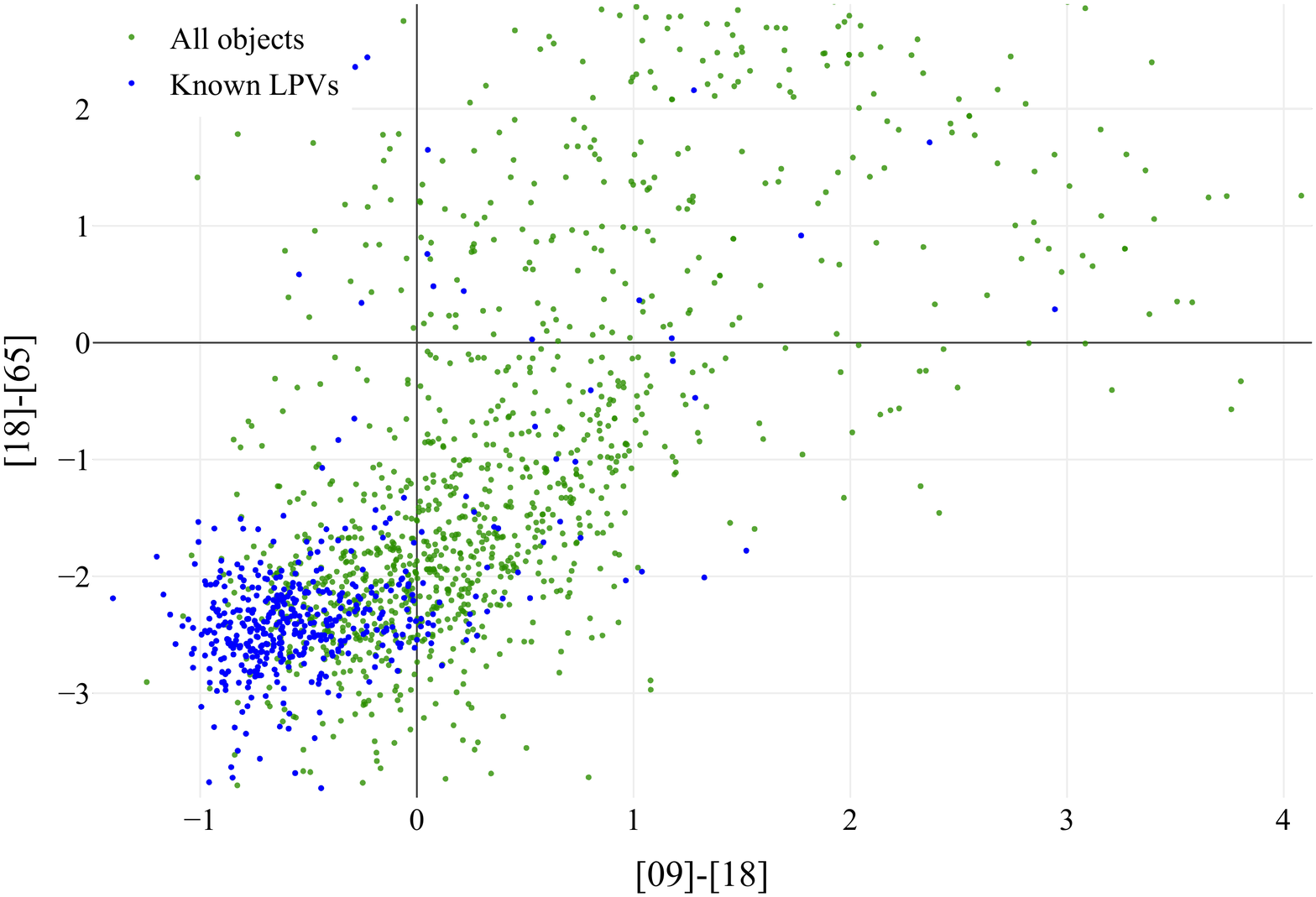} 
% \vspace*{-1.0 cm}
 \caption{Akari colour-colour diagram for objects with detected maser emission of H$_2$O, OH or SiO molecules. Blue dots represent known LPVs (Miras and Semiregular variable stars) from GCVS catalog. }
   \label{fig1}
\end{center}
\end{figure}

\begin{figure}
% \vspace*{-2.0 cm}
\begin{center}
 \includegraphics[width=4.5in]{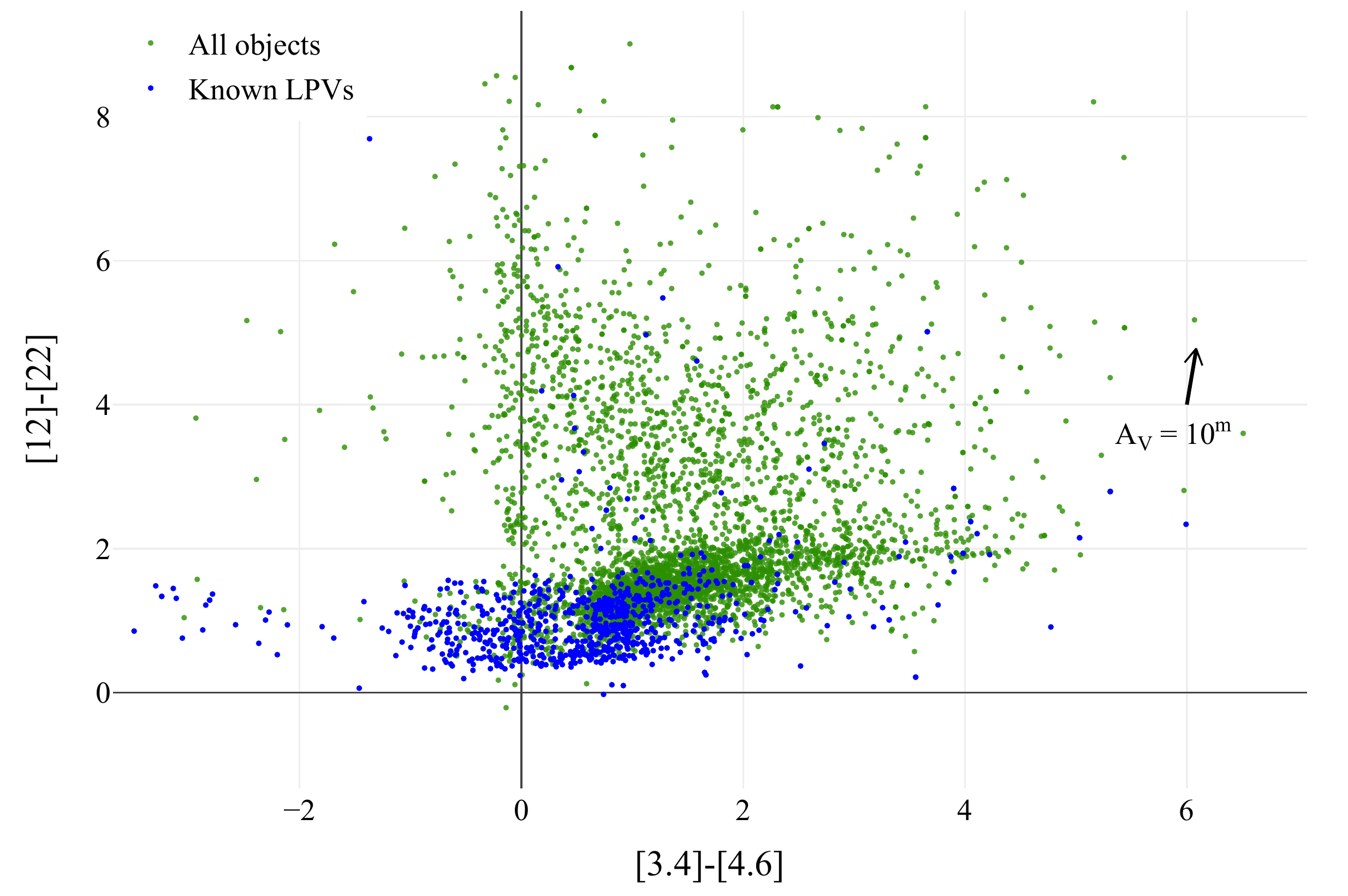} 
% \vspace*{-1.0 cm}
 \caption{WISE colour-colour diagram for objects with detected maser emission of H$_2$O, OH or SiO molecules. Blue dots represent known LPVs (Miras and Semiregular variable stars) from GCVS catalog. Black line represent the reddening vector with A$_V=10^m$}  
   \label{fig1}
\end{center}
\end{figure}

\begin{figure}
% \vspace*{-2.0 cm}
\begin{center}
 \includegraphics[width=4.5in]{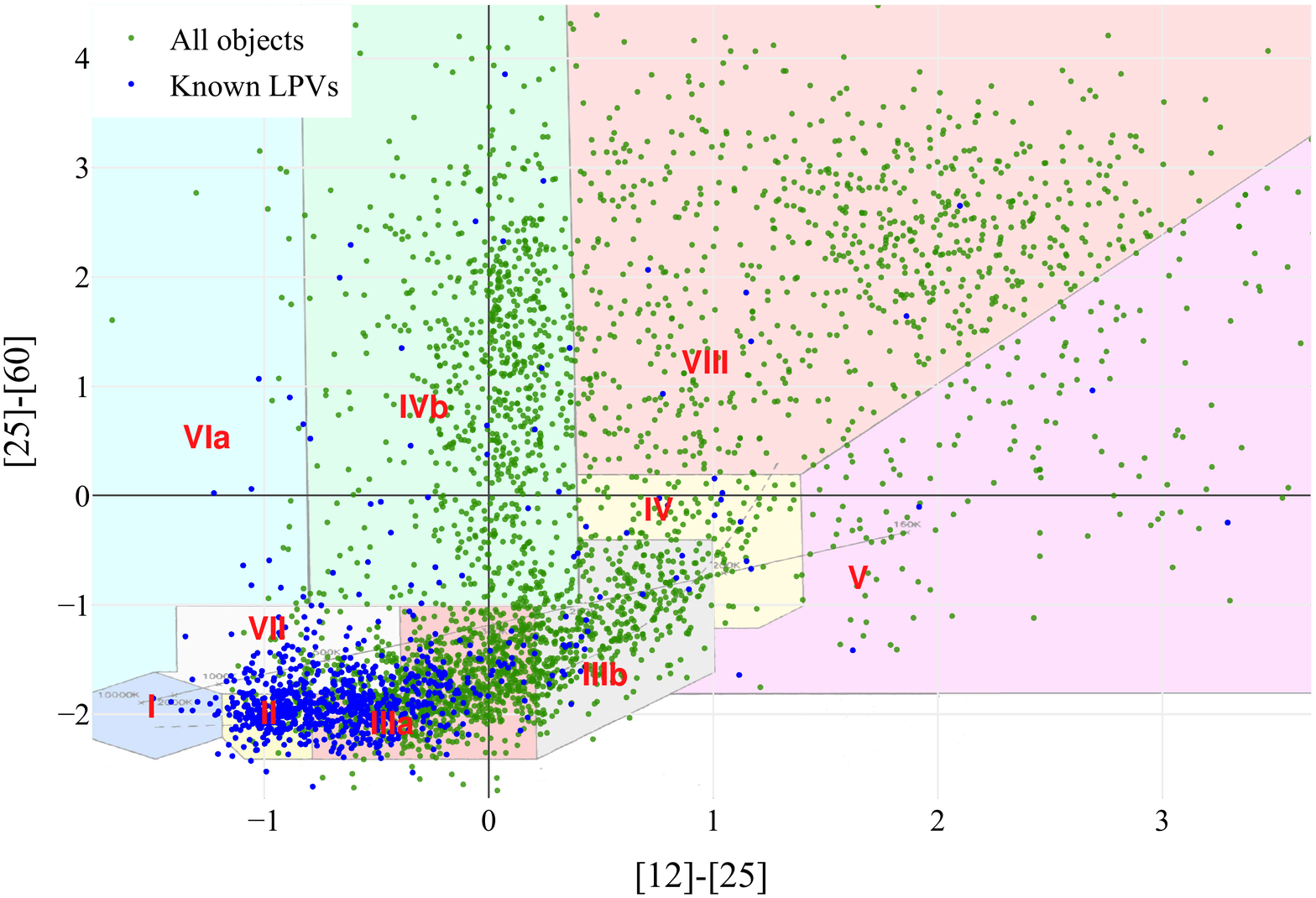} 
% \vspace*{-1.0 cm}
 \caption{IRAS colour-colour diagram for objects with detected maser emission of H$_2$O, OH or SiO molecules. Blue dots represent known LPVs (Miras and Semiregular variable stars) from GCVS catalog. Background image is the regions from \cite{Veen87}.}
   \label{fig1}
\end{center}
\end{figure}

\begin{figure}[b]
% \vspace*{-2.0 cm}
\begin{center}
 \includegraphics[width=4.5in]{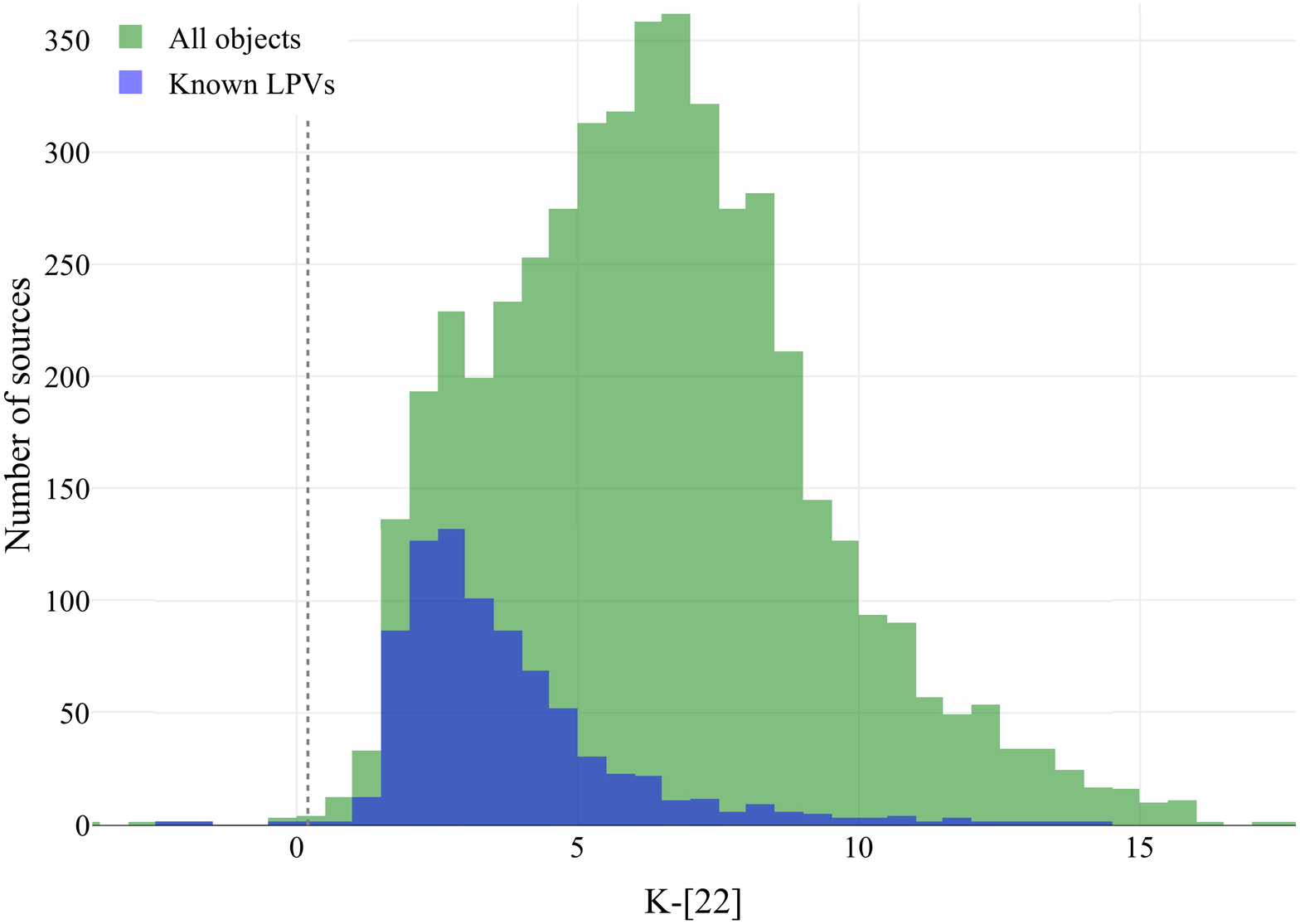} 
% \vspace*{-1.0 cm}
 \caption{Histogram distribution of IR excess as defined in \cite{Wu13}. The green histogram shows objects where maser is detected and blue histograms show known LPVs (Miras and Semiregular variable stars). Vertical dotted line represents the border between the presence of IR excess in objects according to the definition in \cite{Wu13}}
   \label{fig6}
\end{center}
\end{figure}

Analyzing the objects which are not included to the GCVS catalog, we found that 3866 objects from our database have detection in at least one molecule (H$_2$O, OH and SiO) and have no association with GCVS objects. However, star-forming regions may be also included in our sample, because star-forming regions sometimes have color characteristics similar to those of late-type stars (see discussion in \citet{Nakashima15,Nakashima16}). Many authors from the database use color criteria in order to select objects for observations and star-forming regions can be present in the observed source sample. In the figures 2-5 we plot the 2MASS, Akari, WISE and IRAS colour-colour diagrams of selected objects. The known LPVs are selected in these diagrams using the different color. The WISE colour-colour diagram is presented using (W1-W3) and (W2-W4) colours. 2MASS colour-colour diagram is presented using (J-H) and (H-K) colours. The IRAS colour-colour diagram is presented using following colour indices:
$$[12]-[25] = 2.5 \log_{10}(\frac{F_{25}}{F_{12}}), [25]-[60] = 2.5 \log_{10}(\frac{F_{60}}{F_{25}})$$
The Akari colour-colour diagram is presented similar to IRAS but using $[09]-[18]$ and $[18]-[65] $. The Akari diagram is similar to the one in \cite{Yung13} and reveals many objects that exhibit colour characteristics of AGB stars with long-period variability, as defined in \cite{Yung13}. IRAS colour diagrams reveal that most of the  known LPVs fall into groups II, IIIa and VII, according to the definition of \citet{Veen87}. These groups are the following: variable stars with the "young" O-rich circumstellar shells, variable stars with more evolved O-rich circumstellar shells and variable stars with more evolved C-rich circumstellar shells, respectively. All objects with maser detections (green dots) fill other groups of objects in the IRAS colour-colour diagram. 2MASS colour-colour diagram reveals that known LPVs fill quite limited range of the values: $0.5\lesssim(J-H)\lesssim1.5$, $0.2\lesssim(H-K)\lesssim1.3$. Selecting the objects using colour criteria is a possible way to identify the long-period variables, but the more detailed study is necessary in order to find the colour-based criteria for selecting the long-period variable stars.

\section{Infrared excess of selected stars}
\label{sect:data}

Because LPV stars (including M-type) have the  circumstellar shell around, these stars should have IR excess. We follow the scheme described in \cite{Wu13} in order to identify IR excess stars. The authors used WISE and 2MASS photometry in order to select IR-excess stars. The basic criteria is the following: $(K_s-[22])$ should be greater  than some certain value ($\simeq$ 0.2). This constant is a subject for improvement over the years. \cite{Gorlova04,Gorlova06} found that this constant is equal to 0.33. In \cite{Hovhannisyan2009} this value was changed to 0.2. In the paper \cite{Wu13} authors improved the constant value according to the shift between 2MASS and WISE zero-points. The constant value lies in the range 0.26-0.22, depending on the value of $(J-H)$ colour. We used this criterion to check for the presence of the IR excess in our candidates and known Mira-type stars. Our database includes cross-matched WISE and 2MASS photometry of each object, thus it is possible to study the infrared characteristics of the objects. In the Figure \ref{fig6} we plot the histogram of infrared excess for all objects with maser detection. The histogram reveals that most of the objects have IR excess.

Among 832 known LPVs with detected maser emission in at least one molecule only 9 stars (Z Peg, BX Cam, RR Sco, V2584 Oph, AN Dra, V0837 Her, V1677 Aql, V Aqr and TV Aqr)  did not have IR excess, according to the definition  in \cite{Wu13}. All other 823 LPVs have IR excess. 

We considered objects which are not included in the GCVS catalog but have positive maser detection in at least one maser line. We found that only 8 objects (IRAS 16376-4634, IRAS 17320-2734, IRAS 18007-0218, IRAS 18522+0021, IRAS 19126-6941 and IRAS 20084+2750) have no IR excess. The existence of IR excess in the Mira-type candidates confirms the  potential association of these objects with Mira-type variability.

\section{Possible Applications for Variable star studies}
\label{sect:applications}

The maser database presented in this paper has a potential to significantly contribute to research in the field of variable star studies. Firstly, there is a large overlap between the stellar maser sources and variable stars, because a non-negligible amount of stellar maser sources are evolved stars exhibiting pulsation. Secondly, maser properties could be an evolutionary probe of stars in their late stages of evolution. Thirdly, the database system has equipped with functionality allowing cross-identification between different catalogs. With these capabilities, for example, if we cross-check with stellar catalogs at other wavelengths, such as far, mid and near-infrared, ultraviolet, and X-rays, we may be able to identify stars belonging to a key evolutionary stages, such as post-AGB and proto-planetary nebula stages.  

\section{Conclusions}

In this paper, we describe the potential of implementation of astrophysical maser database for the variable star identification and research. The main findings of the paper are the following:

\begin{enumerate}

\item The maser database contains about 11~000 objects with the data on maser emission obtained (including non-detections) for at least one of H$_2$O, OH and SiO molecules. ~4806 objects have maser detection in at least one molecule.

\item Among the known long-period variables in our sample (1803 objects), 46\% of stars (832 objects) have maser emission of at least one molecule (H$_2$O, OH or SiO), thus maser emission may be used as the criterion to identify variable stars.

\item Almost all objects with maser emission have positive IR excess according to the definition in \cite{Wu13}.

\end{enumerate}

\begin{acknowledgements}
The development of web-based software by DAL was funded by Russian Foundation for Basic Research through research project 18-32-00605. AMS was supported by Russian Science Foundation grant 18-12-00193. This work is supported by Act 211 Government of the Russian Federation, agreement No. 02.A03.21.0006.
\end{acknowledgements}

%%% ---------------------------------------------------------------------------------------------------

\bibliographystyle{raa}
\bibliography{bibtex}

%% you can type \apj for ApJ, \aap for A&A, \apss for Ap&SS, etc. Please consult
%% the macro chjaa.cls. You can also find them in aasguide.tex (AASTeX for ApJ, AJ, PASP)
%% Please follow the format of ChJAA's reference list

%\bibitem[Nakashima et al. (2018)]{nak18}
%{Nakashima, J., Engels, D., Hsia, C.-H., Imai, H., Ladeyschikov, D. A., Sobolev, A. M., Yung, B. H. K. \& Zhang, Y.} 2015,
%\textit{IAUS}, 336, in print

\label{lastpage}

\end{document}